\documentclass{ifacconf}

\usepackage{graphicx,xcolor,amsmath,siunitx,amssymb,mathtools}      
\usepackage{natbib,tikz}        

\DeclareMathOperator*{\minimize}{minimize}
\DeclareMathOperator*{\subject_to}{s. t.\: }
\newtheorem{theorem}{Theorem}
\newtheorem{assumption}{Assumption}
\newtheorem{definition}{Definition}
\newtheorem{remark}{Remark}
\newtheorem{lemma}{Lemma}

\usepackage{amsmath,mathtools,tikz,booktabs,bm,multirow,pgfplots}
\usepackage[implicit = false]{hyperref}
\usetikzlibrary{shapes.geometric}
\usepackage{grffile}

\pgfplotsset{compat=newest}
\usetikzlibrary{plotmarks}
\usetikzlibrary{arrows.meta}
\usepgfplotslibrary{patchplots}

\definecolor{mycolor1}{rgb}{0.00000,0.44700,0.74100}%
\definecolor{mycolor2}{rgb}{0.85000,0.32500,0.09800}%
\definecolor{mycolor3}{rgb}{0.46600,0.67400,0.18800}%
\definecolor{dark-gray}{gray}{0.35}
\definecolor{myred}{rgb}{0.6350, 0.0780, 0.1840}
\definecolor{mygreen}{rgb}{0.4660, 0.6740, 0.1880}
\definecolor{myblue}{rgb}{0, 0.4470, 0.7410}

\begin{document}
\begin{frontmatter}

\title{Data-based Moving Horizon Estimation under Irregularly Measured Data\thanksref{footnoteinfo}} 

\thanks[footnoteinfo]{This project has received funding from the European Research Council (ERC) under the European Union’s Horizon 2020 research and innovation programme (grant agreement No 948679).}

\author[IRT]{Tobias M. Wolff} 
\author[IRT]{Isabelle Krauss} 
\author[IRT]{Victor G. Lopez} 
\author[IRT]{Matthias A. Müller}

\address[IRT]{Leibniz University Hannover, Institute of Automatic Control, 30167 Hannover Germany, (e-mail: \{wolff, krauss, lopez, mueller\}@irt.uni-hannover.de).}

\begin{abstract}                
In this work, we introduce a sample- and data-based moving horizon estimation framework for linear systems. We perform state estimation in a sample-based fashion in the sense that we assume to have only few, irregular output measurements available. This setting is encountered in applications where measuring is expensive or time-consuming. Furthermore, the state estimation framework does not rely on a standard mathematical model, but on an implicit system representation based on measured data. We prove sample-based practical robust exponential stability of the proposed estimator under mild assumptions. Furthermore, we apply the proposed scheme to estimate the states of a gastrointestinal tract absorption system.
\end{abstract}

\begin{keyword}
Data-based Estimation, Optimal control, Irregular measurements
\end{keyword}

\end{frontmatter}
\section{Introduction}
\label{sec:introduction}
This work is centered around state estimation, which is needed for, e.g., control, monitoring and fault diagnosis. The most commonly applied state estimation methods are the Kalman filter and the Luenberger observer. Another powerful method is moving horizon estimation (MHE) - an optimization-based state estimation technique \citep{Rawlings2017}. Compared to other state estimation methods, MHE can consider system-inherent constraints directly in the estimation process.


Standard moving horizon estimation is based on the knowledge of an accurate mathematical model. Obtaining such a model can be time-consuming and difficult. Hence, in this work we represent the system dynamics by means of an implicit system representation based on the so-called fundamental lemma \citep{Willems2005}. Using a single persistently exciting trajectory of a controllable system, we can represent any finite-length behavior of that system. This result led to a large body of literature related to data-based control, such as, e.g., predictive control \citep{Berberich2020a,Coulson2019} and state feedback control \citep{DePersis2019}. In recent years, also data-based state estimation received more attention, such as, e.g., data-based unknown input observers \citep{Turan2022}, observers based on the duality principle of control and estimation \citep{Adachi2021}, data-based variants of the Kalman filter \citep{Mishra2024}, and data-based moving horizon estimation frameworks \citep{Wolff2024}.

However, these data-based state estimation methods require online output measurements at every time instant. This might be a restrictive assumption in many applications where measuring at each time instant is expensive, time-consuming, or even impossible. Consider, e.g., biomedical applications in which measuring some plasma hormone concentration implies to take a blood sample in a medical facility which is then analyzed in a separate laboratory. This whole process cannot be repeated too often for cost and time reasons.

In the state estimation context, few works systematically handle irregular output measurements \citep{Krauss2025b,Mishra2024}. \cite{Krauss2025b} introduce a \emph{model-based} MHE scheme for general nonlinear systems that uses irregular output measurements (i.e., a sample-based MHE scheme). The authors exploit the concept of sample-based detectability introduced in \cite{Krauss2025} to show robust stability of the state estimation error. In \cite{Mishra2024}, a data-based state estimator for linear systems was proposed, which leverages the interpolation procedure presented in \cite{Markovsky2022} to recover full-length output sequences.

In contrast to these works, we here extend our data-based MHE framework \citep{Wolff2024} to the case of few and irregular output measurements (Section~\ref{sec:mhe:scheme}) and refer to it as \emph{sample- and data-based} MHE framework. Different from \cite{Krauss2025b}, no knowledge of the mathematical model of the system is required. Furthermore, different from \cite{Mishra2024}, our scheme does not perform any intermediate interpolation step and can also handle the case where no online measurements are available in a certain time window. We prove robust stability of the proposed estimator given a suitable sample-based detectability assumption (Section~\ref{sec:stab:guarantees}). In a simulation study on a biomedical application (Section~\ref{sec:example}), we show that the performance of the sample- and data-based MHE framework can be close to the performance of a benchmark data-based MHE framework albeit using substantially less output measurements. 



\section{Preliminaries and Setup}
\label{sec:preliminaries}
The set of real numbers is denoted by $\mathbb{R}$, the set of integers in the interval $[a,b] \subset \mathbb{R}$ by $\mathbb{I}_{[a,b]}$, and the set of integers greater than or equal to $a \in \mathbb{R}$ by $\mathbb{I}_{\geq a}$. For a symmetric positive definite matrix $P$ and a vector $x = \begin{pmatrix}
	x_1& \dots& x_n
\end{pmatrix}^\top \in \mathbb{R}^n$, we write $||x||_P = \sqrt{x^\top Px}$. We denote the Euclidean norm by $||x||_2$ and the infinity norm by $||x||_\infty$. The maximum (minimum) eigenvalue of a symmetric matrix~$P$ is written as $\lambda_{\max}(P)$ ($\lambda_{\min}(P)$) and the maximum generalized eigenvalue of symmetric and positive definite matrices $P_1$ and $P_2$ as~$\lambda_{\max}(P_1, P_2)$, i.e., the largest scalar $\lambda$ satisfying $\det(P_1-\lambda P_2) = 0$. The identity matrix of dimension $n$ is denoted by $I_n$. A function $\gamma : \mathbb{R}_{\geq 0} \rightarrow \mathbb{R}_{\geq 0} $ is of class $\mathcal{K}$, if $\gamma$ is continuous, strictly increasing, and $\gamma(0) = 0$. A stacked window of a sequence $\{x(k) \}_{k=0}^{N-1}$ from $x(i)$ up to $x(j)$ is denoted by $x_{[i,j]} = \begin{pmatrix} x(i)^\top & \dots & x(j)^\top  \end{pmatrix}^\top$. Finally, the Hankel matrix of depth $L$ of a stacked window ${x}_{[0,N-1]}$ is defined by
\begin{equation*}
	H_L({x}_{[0,N-1]}) = \begin{pmatrix}
		x(0) & x(1) & \dots & x(N-L) \\
		x(1) & x(2) & \dots & x(N-L+1)\\
		\vdots & \vdots & \ddots & \vdots \\
		x(L-1) & x(L) & \dots & x(N-1) \\
	\end{pmatrix} \hspace{-4pt}.
\end{equation*}

Our work heavily relies on the fundamental lemma from \cite{Willems2005}, which holds for linear discrete-time systems of the following form 
\begin{subequations}
	\label{def:system:dynamics}
	\begin{align}
		x(t+1) &= Ax(t) + Bu(t) \label{def:state:transition:function}\\
		y(t) &= Cx(t) + Du(t)
	\end{align}
\end{subequations}
with states $x\in \mathbb{R}^n$, inputs $u \in \mathbb{R}^m$, and outputs $y \in \mathbb{R}^p$. To state the fundamental lemma, we require the definition of persistence of excitation (PE).
\begin{definition}
	\label{def:PE}
	An input sequence $\{u(k)\}_{k=0}^{N-1}$ is persistently exciting of order $L$ if $\mathrm{rank}(H_L(u_{[0,N-1]})) = mL$.
\end{definition}

The fundamental lemma in the classical state-space framework can then be formulated as follows.
\begin{lemma}
	\label{thm:Willems}
	(\cite{Berberich2020}) Suppose \linebreak $u_{[0,N-1]}$, $y_{[0,N-1]}$ is a trajectory of a controllable LTI system (\ref{def:system:dynamics}), where $u_{[0,N-1]}$ is persistently exciting of order~$L+n$. Then, $\bar{u}_{[0,L-1]}$, $\bar{y}_{[0,L-1]}$ is a trajectory of linear system (\ref{def:system:dynamics}) if and only if there exists $\alpha \in \mathbb{R}^{N-L +1}$ such that
	\begin{equation}
		\begin{bmatrix}
			H_L(u_{[0,N-1]}) \\
			H_L(y_{[0,N-1]}) \\
		\end{bmatrix} \alpha = \begin{bmatrix}
			\bar{u}_{[0,L-1]}\\
			\bar{y}_{[0,L-1]} \\
		\end{bmatrix} \label{Willems_Lemma} .
	\end{equation}
\end{lemma}
The lemma states that we can represent \textit{any} finite-length trajectory of a controllable system as linear combinations of windows of one single PE trajectory. 
\begin{remark}
	\label{Minimal_realization}
	In addition to Lemma~\ref{thm:Willems}, we note that the following also holds \cite[Eq. (5)]{Berberich2020}
	\begin{equation}
		\bar{x}_{[0, L-1]} = \sum_{i = 0}^{N-L} \alpha_i x_{[i, L-1 +i]},
	\end{equation}
	where $\bar{x}_{[0, L-1]}$ and $x_{[0, N-1]}$ are state trajectories of system~(\ref{def:system:dynamics}) that correspond to the trajectories $\bar{u}_{[0,L-1]}$, $\bar{y}_{[0,L-1]}$ and $u_{[0,N-1]}$, $y_{[0,N-1]}$ of (\ref{thm:Willems}), respectively, and $\alpha_i$ for $i = 0,\dots, N-L$ correspond to the individual elements of the vector $\alpha$ in (\ref{Willems_Lemma}). This result is important in the context of data-based state estimation as will become clear in Section~\ref{sec:mhe:scheme}.
\end{remark}
Finally, \cite{Willems2005} also prove the following lemma.
\begin{lemma}
	\label{lem:rank}
	Suppose $x_{[0,N-1]}$, $u_{[0,N-1]}$ is an input/state trajectory of the controllable LTI system (\ref{def:system:dynamics}) where $u_{[0,N-1]}$ is persistently exciting of order $L + n$, then the matrix
	\begin{equation}
		\begin{bmatrix}
			H_1(x_{[0,N-L]}) \\
			H_L(u_{[0, N-1]})
		\end{bmatrix}
	\end{equation}
	has full row rank.
\end{lemma}

In this work, we consider two phases. First, an \textit{offline phase}, in which we have $N$ noisy measurements of the outputs \textit{and} states at every time instant available, i.e., 
\begin{align}
	\tilde{x}^d(t) &= x^d(t) + \varepsilon^d_x(t) \label{def:offline:states},\\
	\tilde{y}^d(t) &= y^d(t) + \varepsilon_y^d(t) \label{def:offline:outputs},
\end{align}
where the noise-free state and output measurements are denoted by $x^d$ and $y^d$, respectively. The offline state and output measurement noise are denoted by $\varepsilon^d_x$ and $\varepsilon_y^d$, respectively, and we assume that $ ||\varepsilon_x^d(t)||_\infty  \leq \bar{\varepsilon}_x^d$ and $||\varepsilon_y^d(t)||_\infty \leq \bar{\varepsilon}_y^d$. Additionally, we define
\begin{equation}
	\label{def:varepsilon:max}
	\bar{\varepsilon}^d = \max \{\bar{\varepsilon}_x^d, \bar{\varepsilon}_y^d\}.
\end{equation}
The assumption of having regular state measurements available in an offline phase is standard in the data-based state estimation literature \citep{Turan2022,Mishra2024,Wolff2024} and is needed to estimate system states in a physically meaningful realization. This setting is also fulfilled in various applications, where in a laboratory setting during the offline phase, one can equip the system with additional sensors that might be too expensive or impossible to employ during later runtime of the system (i.e., during the online phase). For instance, this setting is encountered in the automotive industry, where additional sensors are installed into a vehicle in the development stage (i.e., the offline phase). However, in a series production (i.e., the online phase), only a limited amount of sensors is installed in the vehicle and instead observers are employed to get state information. Having state measurements in an offline phase is crucial for the results of our work, since, as will become clear in the next section, the online state estimate will result from linear combinations of the offline state measurements. This guarantees that the online state estimates are in a physically meaningful realization. Note that without the offline measured state trajectory, the realizations of the estimated states would be unknown, since different state trajectories can explain the given input/output data as it is well known, e.g., from the context of subspace identification \citep{VanOverschee1997}. 


In the \textit{online phase}, we only have noisy measurements of the outputs, i.e., 
\begin{align}
	y(t) = Cx(t) + Du(t) + v(t) \label{def:online:outputs}
\end{align}
with $v(t) \in \mathbb{V} \subset \mathbb{R}^p$ but \textit{not} of the states available. Additionally, we allow for only \textit{few and irregular} output measurements, which is encountered in various applications where measuring is time-consuming and/or expensive. 

\section{Sample- and Data-based MHE Framework}
\label{sec:mhe:scheme}
Before explaining the sample- and data-based MHE framework in detail, we first need to introduce some notation. Let $\Omega = \{t_i\}_{i=1}^\infty$ with $0\leq  t_i < t_{i+1}$ be the \emph{infinite} set of time instances at which output measurements will be taken during the execution of the state estimation scheme.  Note that this set is used for analysis only and does not need to be fixed before the application of the estimator. Furthermore, we require a reduced Hankel matrix of the output measurements that is composed of the block rows that correspond to online measurement time instances, i.e., we introduce $H_{L_t}^{\tilde{\Omega}_t}(\tilde{y}^d_{[0,N-2]})$, with $L_t \coloneqq \min\{t,L\}$, which is a submatrix of $H_{L_t}(\tilde{y}^d_{[0,N-2]})$. To specify the row indices of $H_{L_t}^{\tilde{\Omega}_t}$, it will be useful to define the set of ``measurement block row indices" $\tilde{\Omega}_t \coloneqq \{ j \in \mathbb{N}| j = i - (t-L_t) +1, i \in \Omega\cap [t-L_t, t-1]\}$. Then, $H_{L_t}^{\tilde{\Omega}_t}$ is formed by the block rows of $H_{L_t}$ corresponding to the indices in $\tilde{\Omega}_t$.
By $\bar{x}_{[-L ,0]}(t) \coloneqq \begin{pmatrix}
	\bar{x}(t-L|t)^\top   & \dots & \bar{x}(t|t)^\top
\end{pmatrix}^\top$, we denote the estimated state sequence from time $t-L_t$ up to time~$t$, estimated at time $t$ (and we use a similar notation for the slacks $\sigma^x$ and $\sigma^y$, the role of which is explained below). The measured input sequence from time~$t-L_t$ up to time $t-1$ is denoted by $u_{[t-L_t,t-1]}$. By $\tilde{y}_{[t-L_t,t-1] \cap \Omega}$, we denote the vectors whose entries are the noisy output measurements that are in the current horizon from $t-L_t$ up to $t-1$. Finally, we write $\sigma^y_{[-L_t,-1] \cap \Omega}(t)$ to denote\footnote{With a slight abuse of notation, the subindex $[-L_t,-1] \cap \Omega$ here refers to the time indices in  $[t-L_t,t-1] \cap \Omega$.} the estimated measurement noise at the online measurement time instances.

We are now ready to introduce the sample- and data-based MHE framework, which is based on the formulation proposed in our previous work \citep{Wolff2024} and extended here to the case of irregular measurements. At each time $t$, given a horizon length $L$, and input/output measurements $u_{[t-L_t,t-1]}$, $\tilde{y}_{[t-L_t,t-1] \cap \Omega}$, solve
\begin{subequations} 
	\label{MHE_noisy}
	\begin{align}
		&\minimize_{\substack{\bar{x}_{[-L_t,0]}(t), \alpha(t),\\  \sigma_{[-L_t,-1] \cap \Omega}^y (t), \sigma_{[-L_t,0]}^x (t)}} \nonumber \\ &\hspace{.4cm}J(\bar{x}(t-L_t|t),\sigma^y_{[-L_t,-1] \cap \Omega}(t),\sigma^x_{[-L_t, 0]}(t), \alpha(t))		 \label{cost_function_noisy} \\
		&\subject_to \hspace{0.2cm} \begin{bmatrix}
			H_{L_t}(u^d_{[0,N-2]}) \\
			H_{L_t}^{\tilde{\Omega}_t}(\tilde{y}^d_{[0,N-2]}) \\
			H_{L_t+1}(\tilde{x}^d_{[0,N-1]}) \\
		\end{bmatrix}
		\alpha(t) \nonumber \\
		& \hspace{2cm}=\begin{bmatrix}
			u_{[t-L_t,t-1]}\\
			\tilde{y}_{[t-L_t,t-1] \cap \Omega} -\sigma_{[-L_t,-1] \cap \Omega}^y (t)\\
			\bar{x}_{[-L_t ,0]}(t) + \sigma_{[-L_t,0]}^x (t)\\
		\end{bmatrix} \label{System_Dynamics_noisy}\\
		& \hspace{2.5cm} \bar{x}_{[-L_t,0]}(t) \in \mathbb{X}
		\label{state_constraint_noisy}
	\end{align}
	with
	\begin{align}
		&J(\bar{x}(t-L_t|t),\sigma^y_{[-L_t,-1] \cap\Omega}(t), \sigma^x_{[-L_t, 0]}(t), \alpha(t)) \nonumber \\ 
		&\coloneqq  \Gamma(\bar{x}(t-L_t|t))+ \sum_{k \in \mathbb{I}_{[t-L_t,t-1]} \cap \Omega}^{} l_k(\sigma^y(k|t)) \nonumber \\
		&+ c_{\sigma^x}||\sigma^x_{[-L_t, 0]}(t)||^2+ c_\alpha\big( (\bar{\varepsilon}_x^d)^2+ (\bar{\varepsilon}_y^d)^2\big)||\alpha(t)||^2	.	
		\label{MHE_noisy_cost}
	\end{align}
\end{subequations}
In the cost function~(\ref{MHE_noisy_cost}), we consider a prior weighting defined as follows
\begin{align}
	\Gamma(\bar{x}(t-L_t|t)) \coloneqq 2||\bar{x}(t-L_t|t) -\hat{x}(t-L_t)||_{P_2}^2\eta^{L_t},  
\end{align}
with a positive definite weighting matrix $P_2$, the state estimate at time $t-L_t$, i.e., $\hat{x}(t-L_t)$, and a discount factor $\eta\in [0,1)$, which will be crucial to prove robust stability as shown in Section~\ref{sec:stab:guarantees}. This prior weighting penalizes the deviation from the initial element of the estimated state sequence to the state estimate at time $t-L_t$. This type of prior weighting is called \textit{filtering} prior in the MHE literature and has beneficial theoretical properties \citep{Allan2019}. In the second term of (\ref{MHE_noisy_cost}), we penalize the estimated measurement noise $\sigma^y$ at the online measurement time instances, where 
\begin{align}
	l_k(\sigma^y(k|t)) \coloneqq ||\sigma^y(k|t)||_R^2\eta^{t-k-1}
\end{align}
with a positive semi-definite $R$. The third and fourth term correspond to regularization terms. The term related to~$\sigma^x$ brings the estimated state sequence closer to the span of the offline state measurements. The term related to~$\alpha$ limits the amplification of the noise affecting the offline state and output measurements measurement, compare also the discussion in \cite{Wolff2024}. 

In the constraints~(\ref{System_Dynamics_noisy}), we consider the span of the offline measured input/state/output trajectory. Note that the noisy online output measurements~$\tilde{y}$ might not necessarily be in the span of the noisy offline measurements. This is why we need to consider the estimated measurement noise~$\sigma^y$ on the right-hand side of the constraints (\ref{System_Dynamics_noisy}).

Furthermore, the offline measured state trajectory is also noisy, which requires the introduction of the slack $\sigma^x$ in the last row of the constraints (\ref{System_Dynamics_noisy}). Otherwise, we cannot guarantee that the estimated state trajectory satisfies the state constraints~(\ref{state_constraint_noisy}). Here, $\mathbb{X}$ in (\ref{state_constraint_noisy}) is some known constraint set that contains all possible system states and typically corresponds to system-inherent constraints, such as nonnegativity constraints of chemical concentrations. 

The solutions to the optimization problem~(\ref{MHE_noisy}) are denoted by $\hat{x}_{[-L_t,0]}(t)$, $\hat{\alpha}(t)$, $\hat{\sigma}^y_{[-L_t, -1] \cap \Omega}(t) $, and $\hat{\sigma}^x_{[-L_t, 0]}(t)$. The state estimate at time $t$ is set to the last element of the estimated sequence, i.e., $\hat{x}(t) \coloneqq \hat{x}(t|t)$.

\section{Robust Stability Guarantees}
\label{sec:stab:guarantees}
In this section, we develop robust stability guarantees of the sample- and data-based MHE scheme. Before showing the main theorem, we start by introducing some assumptions and definitions. We are interested in showing robust stability according to the following definition.
\begin{definition}
	\label{def:robust:stability}
	Consider system (\ref{def:system:dynamics}) subject to disturbances $v \in \mathbb{V}$ in the online phase, compare (\ref{def:online:outputs}), and $\varepsilon_x$, $\varepsilon_y$ in the offline phase, compare (\ref{def:offline:states}) - (\ref{def:varepsilon:max}), respectively. A state estimator is sample-based practically robustly exponentially stable (pRES) if there exist a function $\gamma \in \mathcal{K}$, constants $c_1, c_2 \geq 0$, and $\lambda_{1}, \lambda_{2} \in [0,1)$ such that for all $x(0)$, $\hat{x}(0) \in \mathbb{X}$, and all $v \in \mathbb{V}$ the following is satisfied for all $t \in \mathbb{I}_{\geq 0}$:
	\begin{align}
		||x(t) - \hat{x}(t)&|| \nonumber \leq  c_{1} ||x(0) - \hat{x}(0)|| \lambda_{1}^t \\ 
		&+ \sum_{\tau \in \mathbb{I}_{[0, t-1]}\cap \Omega}^{} c_{2} ||v(t-\tau)|| \lambda_{2}^\tau + \gamma( \bar{\varepsilon}^d).  \label{eq:pRGES}
	\end{align}
\end{definition}
In the employed detectability notion (see Definition~\ref{def:sample-based:IOSS} below), we rely on a set $K$ defined as follows.
\begin{definition}
	\label{def:sampling:set}
	\cite[Def. 4]{Krauss2025}
	Consider an infinitely long sequence $\Delta = \{\delta_1, \delta_2, \dots \}$ with $\delta_i \in \mathbb{I}_{\geq 0}$, $i \in \mathbb{I}_{>0}$ and $\max_i \delta_i \eqcolon \delta_{\max} <\infty$. The set $K_i = \{t_1^i, t_2^i, \dots \}$ refers to an infinite set of time instances defined as
	\begin{align*}
		t_1^i = \delta_i, \: \: t_2^i = t_1^i + \delta_{i+1}, \: \: \dots, \: \: t_j^i =  t_{j-1}^i + \delta_{i+j-1}, \: \: \dots 
	\end{align*}  
	The set $K$ then refers to a set of sets containing all $K_i$, $i \in \mathbb{I}_{\geq 0}$.
\end{definition}

The set $\Delta$ defines a pattern for taking samples at certain time instances. The set $K_i$ corresponds to one such possible sampling scheme that starts at $\delta_i$, and $K$ is a set of sets containing all the possible sampling schemes $K_i$ compatible with $\Delta$.
This definition is useful to analyze detectability uniformly in time. 
We refer the interested reader to \cite{Krauss2025} for a more detailed discussion regarding Definition~\ref{def:sampling:set}.

Next, we define our crucial detectability notion, namely sample-based uniform exponential incremental output-to-state stability (i-OSS). In this definition we use $h(x, u) \coloneqq Cx + Du$.
\begin{definition}
	\label{def:sample-based:IOSS}
	System (\ref{def:system:dynamics}) is sample-based uniformly exponentially i-OSS with respect to the sampling set $K$ if there exist $P_1$, $P_2\succ 0$, $R \succeq 0$ and $\eta \in [0,1)$ such that for any pair of initial conditions $\tilde{x}(0), \tilde{x}'(0) \in \mathbb{X}$ and for all input sequences $\{u(k)\}_{k=0}^\infty$ it holds for all $t \geq 0$ and any $K_i \in K$ that 
	\begin{align}
		&||\tilde{x}(t) - \tilde{x}'(t)||_{P_1}^2 \leq ||\tilde{x}(0) - \tilde{x}'(0)||_{P_2}^2 \eta^t \nonumber\\ 
		&+\sum_{j \in \mathbb{I}_{[0,t-1]} \cap K_i} \eta^{t-j-1} ||h(\tilde{x}(j), u(j)) - h(\tilde{x}'(j), u(j))||_R^2. 
	\end{align}
\end{definition}
This notion corresponds to a sample-based and exponential version of incremental input/output-to-state stability (i-IOSS), which has been used frequently in the MHE literature to prove stability of various MHE schemes \citep{Knuefer2018,Allan2021a,Hu2023}. In fact, i-IOSS is necessary and sufficient for the existence of a robustly stable state estimator \citep{Knuefer2020,Allan2021,Knuefer2023}. Here, we consider a sample-based and exponential variant of the standard i-IOSS notion and do not consider inputs (i.e., process noise in (\ref{def:state:transition:function})), since the fundamental lemma is defined for systems without process noise, compare the discussion in \cite[Rmk. 4]{Wolff2024} for possible extensions to the case of additional process noise. Here, the ``uniformity" notion is employed since the right-hand side in Definition \ref{def:sample-based:IOSS} holds uniformly for all control input sequences. 

To prove robust stability, we need the following detectability assumption.
\begin{assumption}
	\label{ass:sample-based:IOSS}
	Consider a set $K$ as in Definition~\ref{def:sampling:set}. System~(\ref{def:system:dynamics}) is sample-based exponential i-OSS according to Definition~\ref{def:sample-based:IOSS} for this $K$. Moreover, for the actual set of measurement time instances $\Omega$, it holds that $\Omega \in K$.
\end{assumption}
Conditions for Assumption~\ref{ass:sample-based:IOSS} to hold for linear systems (\ref{def:system:dynamics}) are investigated in \cite{Krauss2025b}. 

Furthermore, we need to ensure that we collect sufficiently rich data in the offline phase as stated in the following assumption.
\begin{assumption}
	\label{ass:persistence:of:excitiation}
	The considered system~(\ref{def:system:dynamics}) is controllable and the offline input $\{u^d(k)\}_{k=0}^{N-1}$ applied to the system is persistently exciting of order $L+n+1$.
\end{assumption}
Next, we require that the states and inputs evolve in compact sets.
\begin{assumption}
	\label{ass:compact:sets}
	The states $x$ and control inputs $u$ of system~(\ref{def:system:dynamics}) evolve in compact sets $\mathbb{U}$ and $\mathbb{X}$, i.e., $u(t) \in \mathbb{U}$ and $x(t)\in \mathbb{X}$ for all $t \in \mathbb{I}_{\geq 0}$, respectively. 
\end{assumption}
If necessary, one can apply a pre-stabilizing controller to guarantee that the states of the system do not grow unboundedly. Furthermore, in most cases, the control inputs have certain actuator limits, which means that one can easily find a corresponding set $\mathbb{U}$.

We are now ready to state the main theorem of this work. In this theorem, we give sufficient conditions for the existence of a minimal horizon length $L_{\min}$ such that the proposed estimator is sample-based pRES. 
\begin{theorem}
	\label{thm:robust:stability:sample:based:MHE}
	Consider that noisy offline data as specified in~(\ref{def:offline:states}) and (\ref{def:offline:outputs}) and noisy online data as defined in (\ref{def:online:outputs}) are available to the sample- and data-based MHE framework~(\ref{MHE_noisy}). Let Assumptions~\ref{ass:sample-based:IOSS} - \ref{ass:compact:sets} hold and the horizon length $L \in \mathbb{I}_{\geq \delta_{\max}}$ be chosen such that $ 16\lambda_{\max}(P_2,P_1)^2 \eta^{L} <1$ is satisfied. Moreover, let $c_\alpha$ and $c_{\sigma^x}$ in (\ref{MHE_noisy_cost}) be such that 
	\begin{align}
		c_\alpha &\geq \max\bigg\{ \Big(2\lambda_{\max} (P_2) +\frac{\lambda_{\max}(P_1)}{\lambda_{\max}(P_2,P_1)}\Big)nN,  \nonumber \\	
		&\hspace{1.5cm}2\frac{\eta - \eta^{L}}{1- \eta} p N  \frac{\lambda_{\max}(R)}{\lambda_{\max}(P_2,P_1)}\bigg\} \label{def:c_alpha}\\
		c_{\sigma^x} &\geq \max\bigg\{2\lambda_{\max}(P_2),  \frac{\lambda_{\max}(P_1)}{\lambda_{\max}(P_2,P_1)}\bigg\}. \label{def:c_sigma}
	\end{align}
	Then, there exist $\tilde{\rho} \in [0,1)$ and $\tilde{\gamma} \in \mathcal{K}$ such that the state estimation error satisfies the following bound for all times $t \in \mathbb{I}_{\geq 0}$
	\begin{align}
		&||x(t)- \hat{x}(t)||_{P_1} \leq  4 \tilde{\rho}^{t} \sqrt{\lambda_{\max}(P_2,P_1)} ||x(0) - \hat{x}(0)||_{P_2} 
		\nonumber\\	
		&+  \sqrt{8+4\lambda_{\max}(P_2,P_1)}\sum_{k \in \mathbb{I}_{[0,t -1]} \cap \Omega}^{}||v(k)||_R\tilde{\rho}^{t-k-1} + \tilde{\gamma}(\bar{\varepsilon}^d).
		\label{thm:expression}
	\end{align}
	Hence, the sample- and data-based MHE scheme is sample-based pRES according to Definition~\ref{def:robust:stability}.
\end{theorem}
The proof of the theorem is given in the Appendix. The proof uses ideas from model- and sample-based \citep{Krauss2025b}, and data-based MHE robust stability proofs \citep{Wolff2024}.


\begin{remark}
	\label{rmk:horizon:length}
	In the proof of Theorem~\ref{thm:robust:stability:sample:based:MHE} (in particular in (\ref{def:alpha:max}) and (\ref{def:sigma:max})), it becomes clear that the function $\tilde{\gamma}$ in the error bound (\ref{thm:expression}) increases as the horizon length $L$ increases. This drawback is caused by our proof strategy, where we find bounds for the regularization terms in the cost function that depend on $L$. In practice, it is known that increasing the horizon length does not worsen the estimator performance, and modifying our current proof to reflect this fact is a subject for future work.
\end{remark}

\begin{remark}
	\label{rmk:no:offline:noise}
	If one considers noise-free offline state and output measurements which might be the case if very accurate sensors are used, we can omit the slack $\sigma^x$, since the estimated states will always satisfy the constraints. We also do not need to regularize $\alpha$ in the cost function. The robust stability proof can be modified accordingly by considering these changes and $\bar{\varepsilon}_x^d = \bar{\varepsilon}_y^d = 0$. A standard data-based MHE framework with such a setting is considered in \cite{Wolff2022}. For this setting, the resulting function $\tilde{\gamma}$ in (\ref{thm:expression}) is then \textit{independent} of the horizon length $L$.
\end{remark}

\begin{remark}
	\label{rmk:missing:offline:samples}
	The sample- and data-based MHE framework presented here considers potentially infrequent output measurements in the online phase, but full measurements in the offline phase. In fact, it is possible to consider also irregular offline measurements by leveraging the results presented in \cite{Alsalti2025}. As is shown there, under certain conditions it is possible to construct an implicit data-based system representation only based on trajectories with missing samples. Hence, we could also use this implicit system representation instead of the fundamental lemma to cover this case.
\end{remark}


\section{Application to Gastrointestinal Tract Absorption}
\label{sec:example}
In this section, we apply the sample- and data-based MHE scheme to a biomedical example. In particular, we consider the gastrointestinal tract absorption of an oral medication intake from \cite{Mak1978} with adapted parameters to better visualize the results. As medication, we consider \textit{levothyroxine} (L-$T_4$) that is needed in case of hypothyroidism (which is a consequence of the autoimmune disease called Hashimoto's thyroiditis) to replace the missing endogenously produced thyroxine. We here consider the following dynamics
\begin{subequations}
	\label{biomedical:example}
	\begin{align}
		\dot{x}_1(t) &= - k_1 x_1(t) +u(t) \\
		\dot{x}_2(t) &= k_1x_1(t) -(k_2 + k_3) x_2(t)\\
		y(t) &= k_3x_2(t)
	\end{align}
\end{subequations}
with $k_1 = 1.3$, $k_2 = 0.15$, and $k_3 = 0.15$. The first state corresponds to the undissolved medication in the gut and the second state to the dissolved medication in the gut. The control input $u$ corresponds to the medication intake. The output measurements are the concentrations of L-$T_4$ in the blood.

We discretize the dynamics exactly with a sampling time of $\delta = 15 \min$ and simulate an offline input/state/output sequence of length $N = 67$. We consider noisy measurements as detailed in (\ref{def:offline:states}) and (\ref{def:offline:outputs}) with $\varepsilon^d_x $ and $\varepsilon^d_y$, respectively, sampled from a truncated normal distribution with zero mean and variance $\sigma_{\varepsilon^d_x}^2= \sigma_{\varepsilon^d_y}^2 = 0.01$. We then choose $\eta = 0.98$, $R = 10^8$ (a large value is necessary to give enough weight to the few measurements that are in the current horizon), $P_2 = I_n$, $c_\alpha = 2\cdot 10^7$ and $c_{\sigma^x} = 2\cdot10^7$, and construct the Hankel matrices as required in (\ref{MHE_noisy}). 

In the online phase, we consider some measurement noise sampled from a truncated normal distribution with zero mean and variance $\sigma^2_v = 0.03$. The horizon length is chosen as $L = 32$. The prior estimate is set to $\hat{x}(0) = \begin{pmatrix}
	0 & 0
\end{pmatrix}^\top$ and the true initial condition to $ x(0) = \begin{pmatrix}
	0.3 & 0.3
\end{pmatrix}^\top$. We simulate the concentrations throughout one day, i.e., $N_{\mathrm{sim}} = 96$. 

\begin{figure}
	\centering
	\includegraphics[scale=0.36]{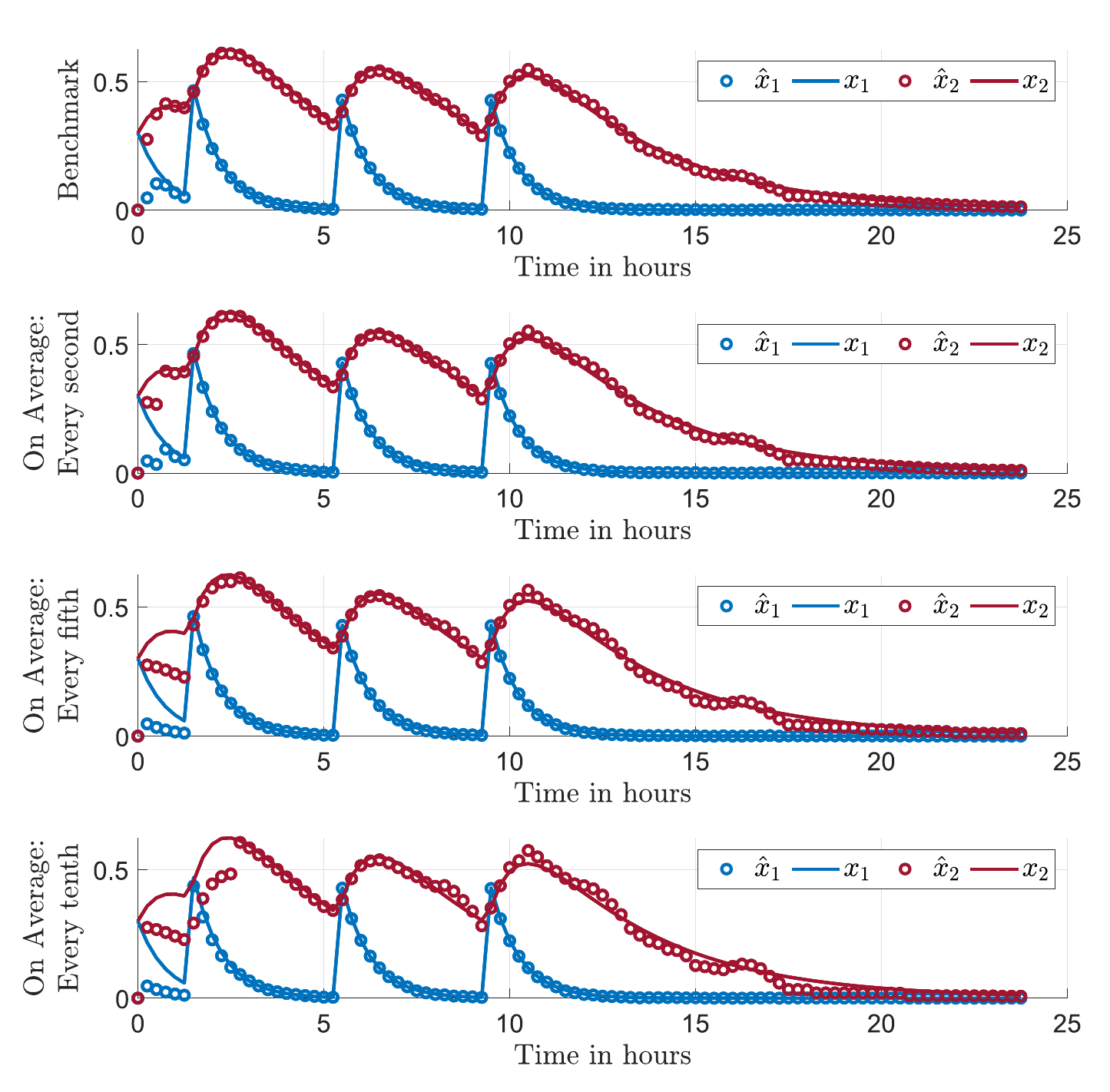}
	\caption{Simulation results of system~(\ref{biomedical:example}) for four different settings. In the first plot, we consider that all output measurements are available to the state estimation scheme (benchmark case). In the following three plots, we consider that the state estimation scheme has only 48, 19, and 9 random measurement available.}
	\label{fig:biomedical:example}
\end{figure}

We first simulate the case, where a complete output sequence (i.e., no missing measurements) is available to the state estimation scheme. This serves as benchmark for the following simulations, where we assume to have only \textit{few and irregular} measurements available. The simulation results are shown in the top plot of Figure~\ref{fig:biomedical:example}. As expected, we observe a very good performance of the proposed MHE framework in the benchmark case.

Next, we perform three different simulations considering irregular measurements. In the first simulation, we assume that 48 random measurements (i.e., on average every second sample) are available to the estimation scheme. In the second and third simulation, we assume that 19 random measurements (i.e., on average every fifth sample) and 9 random measurements (i.e., on average every tenth sample), respectively, are available to the estimation scheme.  The simulation results of all of these simulations are displayed in Figure~\ref{fig:biomedical:example}. 

In addition to the single simulations shown in the plots, we carry out 50 Monte Carlo simulations for each case of available measurements. In the different Monte Carlo simulations, we consider a different initial condition of the true system ($x(0) \sim \mathcal{U}([0,1])$) and a different online noise realization (using the same distribution as introduced above). Then, in each simulation, we compute  the mean squared error (MSE) defined as 
\begin{align}
	\mathrm{MSE} &= \frac{1}{N_{\mathrm{sim}}} \sum_{k= 1}^{N_{\mathrm{sim}}} ||x(k) - \hat{x}(k)||^2.
\end{align}
Finally, we show the averaged MSE (over the 50 Monte Carlo simulations) in Table~\ref{tab:metrics_simulation}.

\begin{table}[t!]
	\centering
	\renewcommand{\arraystretch}{1.3}
	\begin{tabular}{lc} 
		\toprule
		& \multicolumn{1}{c}{averaged MSE} \\ 
		\midrule 
		Benchmark & $7.82 \cdot 10^{-3}$ \\ 
		48 meas. (on average every second) & $1.38 \cdot 10^{-2}$\\
		19 meas. (on average every fifth) & $ 1.88 \cdot 10^{-2}$ \\
		9 meas. (on average every tenth) & $ 3.81 \cdot 10^{-2}$ \\
		\bottomrule
	\end{tabular}
	\vspace{0.2cm}
	\caption{Performance comparison for system~(\ref{biomedical:example}) based on 50 simulations.}
	\label{tab:metrics_simulation}
\end{table}
From Table~\ref{tab:metrics_simulation} and Figure~\ref{fig:biomedical:example}, we obtain the expected result that having all output measurements available results in the best MSE, and that the performance degrades as fewer measurements are taken. In the remaining cases, we first note that the proposed estimator overall performs very well. Second, we conclude that the performance declines only slightly as irregular measurements are taken compared to the benchmark performance. 


\section{Conclusion}
\label{sec:conclusion}
In this work, we introduced a sample- and data-based MHE framework, which exploits the fundamental lemma to obtain a purely data-based system representation. Furthermore, the estimation scheme is designed such that it can handle few and irregular measurements. Finally, we developed robust stability guarantees for the framework. In a biomedical example, we showed that the proposed estimator performs very well.

Our work focuses on the linear case, although MHE has been proven to be very powerful for nonlinear systems. Hence, an extension of our work to nonlinear systems would be interesting. 

{\bibliography{ifacconf}}             

\appendix
\section{Proof of Theorem~\ref{thm:robust:stability:sample:based:MHE}}
\label{sec:proof:theorem}
For a better readability, the proof is divided in three parts. The first part is the construction of a candidate solution for the optimization problem and follows identically from~\cite[Eq. (22) - (27)]{Wolff2024a}. In the second part, we exploit this candidate trajectory together with the optimal solution in the sample-based exponential i-OSS property. In the last part, we combine the sample-based exponential i-OSS property with solutions of the MHE optimization problem to finally prove robust stability.

\textit{Proof:} \textbf{Part I: Construction of a candidate trajectory.}

The candidate trajectory for the optimization problem~(\ref{MHE_noisy}) to be used in the remainder of this proof is described in \cite[Eqs. (22) - (27)]{Wolff2024a} and not shown here for space reasons. Note that we here consider the trajectory from time $t-L_t$ up to time $t$ instead of time $0$ up to time~$t$, as shown in \cite{Wolff2024a}.

\textbf{Part II: Exploitation of the exponential i-OSS property.}

Please note that the first steps of this second part have already been developed in \cite[Thm. 2]{Wolff2024a}. 

In the sample-based exponential i-OSS property, we consider the real unknown  state trajectory parameterized by means of the candidate $\alpha$, compare \cite[Eq. (22)]{Wolff2024a}. This real state trajectory starts at $\tilde{x}(t-L_t) = x(t-L_t)$. Its value at time $t$ corresponds to $\tilde{x}(t) = x(t)$. Furthermore, we consider a state trajectory based on the estimated state sequence. For this, we consider a sequence starting at 
\begin{align}
	&\tilde{x}'(t-L_t) = \nonumber \\ &\hat{x}(t-L_t|t)-H_1(\varepsilon^d_{x,{[0,N-L_t]}})\hat{\alpha}(t) + \hat{\sigma}^x(t-L_t|t) \nonumber \\ 
	&\hspace{1cm}=H_1(x^d_{[0, N-L_t]})\hat{\alpha}(t), \label{def:trajectory:cand:est}
\end{align}
which can be deduced from the last row of (\ref{System_Dynamics_noisy}). After $L_t$ time instances, the state value becomes
\begin{align}
	\tilde{x}'(t) &= \hat{x}(t|t) - H_1(\varepsilon_{x, [L_t-1, N-1]}^d)\hat{\alpha}(t) + \hat{\sigma}^x(t|t) \nonumber\\
	&=  H_1(x^d_{[L_t-1, N-1]}) \hat{\alpha}(t).
\end{align}
Therefore,
\begin{align*}
	||\tilde{x}(t)-&\tilde{x}'(t) ||_{P_1} = \nonumber\\
	&||x(t) - \hat{x}(t) + H_1(\varepsilon^d_{x,{[L_t-1, N-1]}})\hat{\alpha}(t) -\hat{\sigma}^x(t|t)||_{P_1}.
\end{align*} 
Note that $||a|| - ||b|| \leq ||a-b||$ for any vectors $a$ and $b$, which implies
\begin{align*}
	&||x(t) - \hat{x}(t)||_{P_1} - ||-H_1(\varepsilon_{x, [L_t-1, N-1]}^d)\hat{\alpha}(t) +\hat{\sigma}^x(t|t)||_{P_1} \\
	& \leq ||x(t) - \hat{x}(t) + H_1(\varepsilon_{x, [L_t-1, N-1]}^d)\hat{\alpha}(t) -\hat{\sigma}^x(t|t)||_{P_1} \nonumber \\
	&= ||\tilde{x}(t)-\tilde{x}'(t)||_{P_1}.
\end{align*}
Reformulating the above inequality and considering the squared version yields
\begin{align*}
	||x(t) -& \hat{x}(t)||^2_{P_1} \leq  2||\tilde{x}(t) - \tilde{x}'(t)||_{P_1}^2 \nonumber \\
	& + 2||-H_1(\varepsilon_{x, [L_t-1, N-1]}^d)\hat{\alpha}(t) +\hat{\sigma}^x(t|t)||_{P_1}^2,
\end{align*}
where we used the fact that $||a+b||_{P_1}^2 \leq 2||a||_{P_1}^2 + 2||b||_{P_1}^2$ for any vectors $a,b$.

In the following, we use a similar procedure as in \cite{Krauss2025b}. We introduce $\psi_\tau \coloneqq \min \{ i|i\in [\tau, \infty) \cap \Omega\}-\tau$. Note that there is a measurement at time $t-L_t + \psi_{t-L_t}$. Consequently, the following measurements in the interval $[t- L_t + \psi_{t-L_t}, t-1]$ correspond to the pattern given in Definition~\ref{def:sampling:set} using a contiguous subsequence of~$\Delta$.
Then, we can replace the difference $||\tilde{x}(t) - \tilde{x}'(t)||^2_{P_1}$ by means of the sample-based exponential i-OSS condition (recall Definition~\ref{def:sample-based:IOSS}, Assumption~\ref{ass:sample-based:IOSS}, and (\ref{def:trajectory:cand:est})) and obtain
\begin{align}
	&||x(t) - \hat{x}(t)||^2_{P_1} \nonumber \\
	&\leq  2||\tilde{x}(t-L_t+\psi_{t-L_t})  - \tilde{x}'(t-L_t+\psi_{t-L_t})||_{P_2}^2 \eta^{L_t-\psi_{t-L_t}} \nonumber\\ 
	&+2\sum_{k \in \mathbb{I}_{[t-L_t+\psi_{t-L_t},t-1]} \cap \Omega} \eta^{t-k-1} ||h(\tilde{x}(k), u(k)) \nonumber\\
	&\hspace{1cm}- h(\tilde{x}'(k), u(k))||_R^2 \nonumber \\
	& + 2||-H_1(\varepsilon_{x, [L_t-1, N-1]}^d)\hat{\alpha}(t) +\hat{\sigma}^x(t|t)||_{P_1}^2. \label{eq:eOSS:sample1}
\end{align}
Since $\psi_{t-L_t} \leq \delta_{\max}$, and since there are no measurements in the interval $[t-L_t,t-L_t+\psi_{t-L_t}-1]$, it follows from Assumption~\ref{ass:sample-based:IOSS} and the properties of the generalized eigenvalues that
\begin{align}
	&||\tilde{x}(t-L_t+\psi_{t-L_t})  - \tilde{x}'(t-L_t+\psi_{t-L_t})||_{P_2}^2 \nonumber \\
	&\leq \lambda_{\max}(P_2,P_1)||\tilde{x}(t-L_t+\psi_{t-L_t})  \nonumber \\
	&\hspace{1cm }- \tilde{x}'(t-L_t+\psi_{t-L_t})||_{P_1}^2 \nonumber \\
	& \leq  \lambda_{\max}(P_2,P_1) \eta^{\psi_{t-L_t}}||\tilde{x}(t-L_t)  - \tilde{x}'(t-L_t)||_{P_2}^2 \label{eq:aux:sample:expression}.
\end{align}

Substituting (\ref{eq:aux:sample:expression}) in (\ref{eq:eOSS:sample1}) results in
\begin{align*}
	&||x(t) - \hat{x}(t)||^2_{P_1} \nonumber \\
	&\leq  2\lambda_{\max}(P_2,P_1) \eta^{L_t}||\tilde{x}(t-L_t)  - \tilde{x}'(t-L_t)||_{P_2}^2 \nonumber\\ 
	&+2\sum_{k \in \mathbb{I}_{[t-L_t,t-1]} \cap \Omega} \eta^{t-k-1} ||h(\tilde{x}(k), u(k)) \nonumber\\
	&\hspace{1cm}- h(\tilde{x}'(k), u(k))||_R^2 \nonumber \\
	& + 2||-H_1(\varepsilon_{x, [L_t-1, N-1]}^d)\hat{\alpha}(t) +\hat{\sigma}^x(t|t)||_{P_1}^2. 
\end{align*}
We replace $\tilde{x}$ and $\tilde{x}'$ in the first term as explained at the beginning of Part~II above and obtain
\begin{align}
	&||x(t) - \hat{x}(t)||^2_{P_1} \nonumber \\
	&\leq  2\lambda_{\max}(P_2,P_1) \eta^{L_t}||x(t-L_t)  - \hat{x}(t-L_t|t)\nonumber\\
	&\hspace{1cm}+ H_1(\varepsilon^d_{x, [0, N-L_t]})\hat{\alpha}(t) - \hat{\sigma}^x(t-L_t|t)||_{P_2}^2 \nonumber\\ 
	&+2\sum_{k \in \mathbb{I}_{[t-L_t,t-1]} \cap \Omega} \eta^{t-k-1} ||h(\tilde{x}(k), u(k)) \nonumber\\
	&\hspace{1cm}- h(\tilde{x}'(k), u(k))||_R^2 \nonumber \\
	& + 2||-H_1(\varepsilon_{x, [L_t-1, N-1]}^d)\hat{\alpha}(t) +\hat{\sigma}^x(t|t)||_{P_1}^2. \label{eq:eUOSS_noisy1}
\end{align}
In order to obtain an expression for $h(\tilde{x}'(k), u(k))$, from the second block row of (\ref{System_Dynamics_noisy}), we establish
\begin{align}
	&H_{L_t}^{\tilde{\Omega}_t}(y^d_{[0,N-2]})\hat{\alpha}(t)  +H_{L_t}^{\tilde{\Omega}_t}(\varepsilon_{y,[0, N-2]}^d)\hat{\alpha}(t) \nonumber \\ 	
	&\hspace{2cm}=\tilde{y}_{[t-L_t,t-1] \cap \Omega} -\hat{\sigma}_{[-L_t,-1] \cap \Omega}^y (t). \label{eq:output_eUOSS}
\end{align}
Since for $k\in \mathbb{I}_{[t-L_t,t-1]}\cap\Omega$, it holds that $h(\tilde{x}'(k),u(k)) =H_1(y^d_{[k-(t-L_t),k-t+N]}) \hat{\alpha}(t) $, from (\ref{eq:output_eUOSS}) it follows that 
\begin{align*}
	&\sum_{k\in \mathbb{I}_{[t-L_t, t-1]} \cap\Omega }^{}h(\tilde{x}'(t), u(k)) \nonumber \\ 
	&=\sum_{k\in \mathbb{I}_{[t-L_t, t-1]} \cap \Omega}^{}\big(\tilde{y}(k) - \hat{\sigma}^y(k|t) \nonumber \\ 
	& - H_1(\varepsilon_{y,[k-(t-L_t),k-t+N]}^d)\hat{\alpha}(t)\big) .
\end{align*}
Using this result, adding zero to the first term on the right hand side of (\ref{eq:eUOSS_noisy1}), and repeatedly using the fact that $||a+b||^2_P \leq 2||a||^2_P + 2||b||^2_P$ for $P \succ 0$, we can rewrite (\ref{eq:eUOSS_noisy1}) as
\begin{align}
	&||x(t)- \hat{x}(t)||_{P_1}^2   \nonumber\\
	&\leq  8\lambda_{\max}(P_2,P_1)||x(t-L_t) - \hat{x}(t-L_t)||_{P_2}^2 \eta^{L_t} \nonumber\\	
	&+ \sum_{k\in \mathbb{I}_{[t-L_t, t-1]} \cap \Omega}^{} \Big(8||v(k)||_R^2\eta^{t-k-1} +4|| \hat{\sigma}^y(k|t)||_R^2\eta^{t-k-1} \nonumber\\	
	&+ 8||H_1(\varepsilon_{y,[k-(t-L_t),k-t+N]}^d)\hat{\alpha}(t)||_R^2 \eta^{t-k-1}\Big)  \nonumber\\	
	&+8\lambda_{\max}(P_2,P_1)||\hat{x}(t-L_t) - \hat{x}(t-L_t|t) ||_{P_2}^2\eta^{L_t}\nonumber\\	
	&+8\lambda_{\max}(P_2,P_1)||H_1(\varepsilon^d_{x, [0, N-L_t]})\hat{\alpha}(t)||^2_{P_2}\eta^{L_t} \nonumber \\ &+8\lambda_{\max}(P_2,P_1)||\hat{\sigma}^x(t-L_t|t)||_{P_2}^2 \eta^{L_t}\nonumber \\
	&+ 4||H_1(\varepsilon_{x, [L_t-1, N-1]}^d)\hat{\alpha}(t)||_{P_1}^2 + 4||\hat{\sigma}^x(t|t)||_{P_1}^2.  \label{e-UOSS_Basic_noisy2}
\end{align}

\textbf{Part III: Combine the exponential i-OSS property with solutions of the MHE optimization problem}

The established bound in~(\ref{e-UOSS_Basic_noisy2}) still depends on terms that come from the optimal solution to the MHE optimization problem~(\ref{MHE_noisy}). In the following, we exploit optimality properties to bound the terms of the optimal solution with terms from the solution of the candidate trajectory. 

Thus, we now bound (\ref{e-UOSS_Basic_noisy2}) by considering the cost of the optimal trajectory, which is given by
\begin{align*}
	J^\ast = &2||\hat{x}(t-L_t|t)- \hat{x}(t-L_t)||_{P_2}^2\eta^{L_t} \nonumber \\
	&+ \sum_{k\in \mathbb{I}_{[t-L_t, t-1]} \cap \Omega}^{}\eta^{t-k-1}||\hat{\sigma}^y(k|t)||_R^2 \nonumber\\	
	&+ c_{\sigma^x}||\hat{\sigma}^x_{[-L_t, 0]}(t)||^2+ c_\alpha( (\bar{\varepsilon}_x^d)^2 + (\bar{\varepsilon}_y^d)^2)||\hat{\alpha}(t)||^2	.
\end{align*}
The first two terms of (\ref{e-UOSS_Basic_noisy2}) do not need to be bounded, since they already consider the real unknown system state and measurement noise. Next, we consider the following terms 
\begin{align}
	&8\lambda_{\max}(P_2,P_1)||\hat{x}(t-L_t) - \hat{x}(t-L_t|t) ||_{P_2}^2\nonumber\\	
	&+4\sum_{k\in \mathbb{I}_{[t-L_t, t-1]} \cap \Omega}^{}|| \hat{\sigma}^y(k|t)||_R^2\eta^{t-k-1}.
\end{align}
These terms appear in the same way in the cost of the optimal trajectory with an additional factor 4 (and $\lambda_{\max}(P_2,P_1)$ for the prior weighting). Next, we consider
\begin{align}
	&\sum_{k\in \mathbb{I}_{[t-L_t, t-1]} \cap \Omega}^{}   8||H_1(\varepsilon_{y,[k-(t-L_t),k-t+N]}^d)\hat{\alpha}(t)||_R^2 \eta^{t-k-1}\nonumber\\	
	&\leq 8\frac{\eta - \eta^{L}}{1- \eta} (\bar{\varepsilon}_y^d)^2 p N \lambda_{\max}(R) ||\hat{\alpha}(t)||^2.
\end{align}
Then, we consider
\begin{align*}
	&8\lambda_{\max}(P_2,P_1)||H_1(\varepsilon^d_{x, [0, N-L_t]})\hat{\alpha}(t)||^2_{P_2}\eta^{L_t} \nonumber \\
	&\hspace{1cm}+ 4||H_1(\varepsilon_{x, [L_t-1, N-1]}^d)\hat{\alpha}(t)||_{P_1}^2 \nonumber\\	
	&\leq (8\lambda_{\max}(P_2,P_1)\lambda_{\max} (P_2) +4\lambda_{\max}(P_1))(\bar{\varepsilon}_x^d)^2nN ||\hat{\alpha}(t)||^2.
\end{align*}
Finally, since $c_\alpha$ and $c_{\sigma^x}$ satisfy (\ref{def:c_alpha}) and (\ref{def:c_sigma}), respectively, we can bound 
the RHS of (\ref{e-UOSS_Basic_noisy2}) as follows 
\begin{align}
	&||x(t)- \hat{x}(t)||_{P_1}^2 \nonumber \\
	&\leq 8\lambda_{\max}(P_2,P_1)||x(t-L_t) - \hat{x}(t-L_t)||_{P_2}^2 \eta^{L_t} \nonumber\\	
	&+ \sum_{k\in \mathbb{I}_{[t-L_t, t-1]} \cap \Omega}^{} \Big(8||v(k)||_R^2\eta^{t-k-1}\Big) + 4 \lambda_{\max}(P_2,P_1) J^\ast 
	\label{e-UOSS_3}
\end{align}
Then, we use 
\begin{align}
	J^\ast \leq J_{\mathrm{cand}} \label{ineq:optimality}
\end{align}
where $J_{\mathrm{cand}}$ is the cost of the candidate trajectory defined by \cite[Eq. (22) - (27)]{Wolff2024a}. Inequality (\ref{ineq:optimality}) holds due to optimality. Considering (\ref{ineq:optimality}) in (\ref{e-UOSS_3}) results in
\begin{align}
	&||x(t)- \hat{x}(t)||_{P_1}^2 \nonumber \\
	&\leq 8\lambda_{\max}(P_2,P_1)||x(t-L_t) - \hat{x}(t-L_t)||_{P_2}^2 \eta^{L_t} \nonumber\\	
	&+ \sum_{k\in \mathbb{I}_{[t-L_t, t-1]} \cap \Omega}^{} \Big(8||v(k)||_R^2\eta^{t-k-1}\Big) \nonumber \\
	&+ 4 \lambda_{\max}(P_2,P_1) \Big(2||x(t-L_t)- \hat{x}(t-L_t)||_{P_2}^2\eta^{L_t} \nonumber \\
	&+ \sum_{k\in \mathbb{I}_{[t-L_t, t-1]} \cap \Omega}^{}\eta^{t-k-1}||v(k)||_R^2 \nonumber\\	
	&+ c_{\sigma^x}||\sigma^x_{[-L_t, 0]}(t)||^2+ c_\alpha( \bar{\varepsilon}_x^d + \bar{\varepsilon}_y^d)||\alpha(t)||^2	\Big).
	\label{e-UOSS_4}
\end{align}
We simplify (\ref{e-UOSS_4}) and obtain
\begin{align}
	&||x(t)- \hat{x}(t)||_{P_1}^2 \nonumber \\
	&\leq  16\lambda_{\max}(P_2,P_1)||x(t-L_t) - \hat{x}(t-L_t)||_{P_2}^2 \eta^{L_t} \nonumber\\	
	&+ (8+4\lambda_{\max}(P_2,P_1))\sum_{k\in \mathbb{I}_{[t-L_t, t-1]} \cap \Omega}^{}||v(k)||_R^2\eta^{t-k-1} \nonumber\\	
	&  + 4\lambda_{\max}(P_2,P_1)c_{\sigma^x}||\sigma^x_{[-L_t, 0]}(t)||^2\nonumber\\
	&+ 4\lambda_{\max}(P_2,P_1)c_\alpha( (\bar{\varepsilon}_x^d)^2 + (\bar{\varepsilon}_y^d)^2)||\alpha(t)||^2		.
\end{align}
We note that $u(t)$ and $x(t)$ evolve in a compact set, compare Assumption~\ref{ass:compact:sets}. Hence, there exist a $u_\mathrm{max}$ and an $x_\mathrm{max}$ so that $  ||u(t)|| \leq u_\mathrm{max}$ and $||x(t)|| \leq x_\mathrm{max} \quad \forall t \in \mathbb{I}_{\geq 0}$. Thus, we can bound $\alpha(t)$ as defined in \cite[Eq. (23)]{Wolff2024a} by
\begin{align}
	||\alpha(t)|| \leq H_{ux}\sqrt{Lu_{\max} + x_{\max}} \eqqcolon \alpha_{\max} \label{def:alpha:max}
\end{align}
with 
\begin{align}
	H_{ux} = \max_{0\leq t \leq L}\left\| \begin{pmatrix}
		H_{L_t}(u^d_{[0, N-1]})\\
		H_1(x^d_{[0, N-L_t-1]})
	\end{pmatrix}^\dagger \right\|
\end{align}
and $\sigma^x$ as defined in \cite[Eq. (25)]{Wolff2024a} by
\begin{align}
	&||\sigma^x_{[-L_t, 0]}(t)||^2\leq n(L_t+1)||H_{L_t+1}(\varepsilon_{x,{[0,N-1]}}^d) ||_{\infty}^2||\alpha_{\max}||^2 \nonumber\\	
	&\leq n(L+1) N^2 (\bar{\varepsilon}_x^d)^2 \alpha_{\max}^2 \eqqcolon \sigma^x_{\max} .\label{def:sigma:max}
\end{align}
We define
\begin{align}
	c(\bar{\varepsilon}^d) &\coloneqq 4\lambda_{\max}(P_2,P_1)c_{\sigma^x}\sigma^{x}_{\max} \nonumber\\
	&+ 4\lambda_{\max}(P_2,P_1)c_\alpha(( \bar{\varepsilon}_x^d)^2 + (\bar{\varepsilon}_y^d)^2)\alpha_{\max}^2.
\end{align}
This results in 
\begin{align*}
	&||x(t)- \hat{x}(t)||_{P_1}^2 \nonumber\\
	& \leq  16\lambda_{\max}(P_2,P_1)||x(t-L_t) - \hat{x}(t-L_t)||_{P_2}^2 \eta^{L_t} \nonumber\\	
	&+ (8+4\lambda_{\max}(P_2,P_1))\sum_{k\in \mathbb{I}_{[t-L_t, t-1]} \cap \Omega}^{} ||v(k)||_R^2\eta^{t-k-1} \nonumber\\
	& \hspace{2cm}+ c(\bar{\varepsilon}^d).
\end{align*}
Now, we set $L_{\min} \in \mathbb{I}$ such that 
\begin{align}
	\rho^{L_{\min}} \coloneqq 16\lambda_{\max}(P_2,P_1)^2 \eta^{L_{\min}} < 1 \label{eq:min}
\end{align}
is satisfied, choose $L\geq L_{\min}$, consider $t\in\mathbb{I}_{\geq L}$ and obtain
\begin{align*}
	&||x(t)- \hat{x}(t)||_{P_1}^2 \leq  \rho^{L}||x(t-L) - \hat{x}(t-L)||_{P_1}^2 \nonumber\\	
	&+ \sum_{k\in \mathbb{I}_{[t-L, t-1]} \cap \Omega}^{} (8+4\lambda_{\max}(P_2,P_1))||v(k)||_R^2\eta^{t-k-1}+ c(\bar{\varepsilon}^d). \label{eq:contraction}
\end{align*}
Applying this contraction recursively and following similar steps as in the proof of \cite[Cor. 1]{Schiller2023}, we get
\begin{align*}
	&||x(t)- \hat{x}(t)||_{P_1} \leq  \tilde{\rho}^{t}\sqrt{\lambda_{\max}(P_2,P_1)}||x(0) - \hat{x}(0)||_{P_2} 
	\nonumber\\	
	&+ \sqrt{8+4\lambda_{\max}(P_2,P_1)}\sum_{k \in \mathbb{I}_{[0,t -1]} \cap \Omega}^{} ||v(k)||_R\tilde{\rho}^{t-k-1} + \tilde{\gamma}(\bar{\varepsilon}^d)
\end{align*}
with $\gamma(\bar{\varepsilon}) \coloneqq (1/(1-\rho)) c(\bar{\varepsilon})$,
$\tilde{\rho} \coloneqq \sqrt{\rho}$, and $\tilde{\gamma} \coloneqq\sqrt{\gamma} \in \mathcal{K}$ resulting in the desired expression. \hfill $\blacksquare$
\end{document}